\documentclass[conference]{IEEEtran}
\IEEEoverridecommandlockouts
\usepackage{cite}
\usepackage{amsmath,amssymb,amsfonts}
\usepackage{algorithmic}
\usepackage{graphicx}
\usepackage{textcomp}
\usepackage{xcolor}
\usepackage{subfigure}
\usepackage{multirow}
\usepackage{url}
\usepackage{booktabs}
\def\BibTeX{{\rm B\kern-.05em{\sc i\kern-.025em b}\kern-.08em
    T\kern-.1667em\lower.7ex\hbox{E}\kern-.125emX}}
\begin{document}

\title{Demystifying the Performance of Bluetooth Mesh: Experimental Evaluation and Optimization}

\author{
\IEEEauthorblockN{Adnan Aijaz, Aleksandar Stanoev, Dominic London, and Victor Marot}
\IEEEauthorblockA{
\text{Bristol Research and Innovation Laboratory, Toshiba Europe Ltd., Bristol, United Kingdom}\\
firstname.lastname@toshiba-bril.com}
}    

\maketitle

\begin{abstract}
Mesh connectivity is  attractive for Internet-of-Things (IoT) applications from various perspectives. The recent Bluetooth mesh specification provides a full-stack mesh networking solution, potentially for thousands of nodes. Although Bluetooth mesh has been adopted for various IoT applications, its performance aspects are not extensively investigated in literature. This paper provides an experimental evaluation of Bluetooth mesh (using Nordic nRF52840 devices) with an emphasis on those aspects which are not well-investigated in literature. Such aspects include evaluation of unicast and group modes, performance under different traffic patterns, impact of message segmentation, and most importantly, latency performance for perfect reliability. The paper also investigates performance enhancement of Bluetooth mesh based on different techniques including parametric adjustments, extended advertisements (introduced in Bluetooth 5.0), power control, and customized relaying. Results provide  insights into system-level performance of Bluetooth mesh while clarifying various important issues identified in recent  studies. 
\end{abstract}

\begin{IEEEkeywords}
Bluetooth, BLE, flooding, IoT, mesh, multi-hop, power control, wireless networks. 
\end{IEEEkeywords}

\section{Introduction}
Mesh networking, which is a key enabler for Internet-of-Things (IoT),  is a promising solution for range extension and improved resilience. Bluetooth is a prominent short-range connectivity technology which has evolved tremendously over the last decade. Bluetooth mesh (released in 2017) \cite{BT_mesh} is the latest addition to the IoT connectivity landscape. The Bluetooth mesh specification provides a full-stack mesh networking solution for IoT. Bluetooth mesh provides a simple and efficient wireless networking solution potentially providing mesh connectivity to thousands\footnote{Bluetooth mesh can support up to 32767 nodes in a network with a maximum of 126 hops.} of nodes without any coordination.  In a concurrent development, the Bluetooth 5.0 standard (released in 2016) \cite{BT_5} introduces various new features for IoT including improved data transmission modes, enhance frequency hopping, and new Physical (PHY) layers.

Bluetooth mesh has been adopted for various IoT applications including home/building automation, smart lighting, and industrial wireless sensor networks. In the smart lighting community, Bluetooth mesh is recognized as the \emph{killer of the lighting switch} due to its distributed control capabilities. Recently, Silvair\footnote{\url{https://silvair-media.s3.amazonaws.com/media/filer_public/c8/cf/c8cfa7fb-278c-450b-89c6-6c5137ca4fc4/25_02_silvair_emc_casestudy.pdf}} has deployed one of the largest Bluetooth mesh networks comprising more than 3500 nodes spread across 17 floors of an office building. 

Despite growing success of Bluetooth mesh in industry, academic studies on its performance aspects portray a mixed picture. One simulation-based study \cite{und_perf_BT_mesh} identifies scalability as its main limitation. Another study following analytic approach \cite{BT_mesh_analysis} claims configuration of wide range of parameters as the main challenge. Yet another study \cite{exp_BT_mesh_1}, conducting experimental evaluation, suggests that Bluetooth mesh is better suited to applications with sporadic traffic patterns. On the other hand, various aspects of Bluetooth mesh performance remain unexplored. One example is the latency performance of Bluetooth mesh as most existing studies are heavily focused toward reliability. Bluetooth mesh stack offers a number of configuration options at different layers that provide the capability of perfect reliability. Hence, the latency associated with perfect reliability requires deeper investigation, particularly under different communication patterns. The role of Bluetooth 5.0 for Bluetooth mesh as well as performance enhancement techniques are not well-investigated. This necessitates further experimental evaluation of Bluetooth mesh and performance insights. Investigating performance trade-offs for Bluetooth mesh is crucial to asses its suitability for new near-real-time and non-real-time IoT applications. This is also important as Bluetooth-based proprietary industrial solutions (e.g., \cite{enc_TII} and \cite{IO_Link_W}) are not compatible with the Bluetooth mesh specification. 

\begin{figure}
\centering
\includegraphics[scale=0.27]{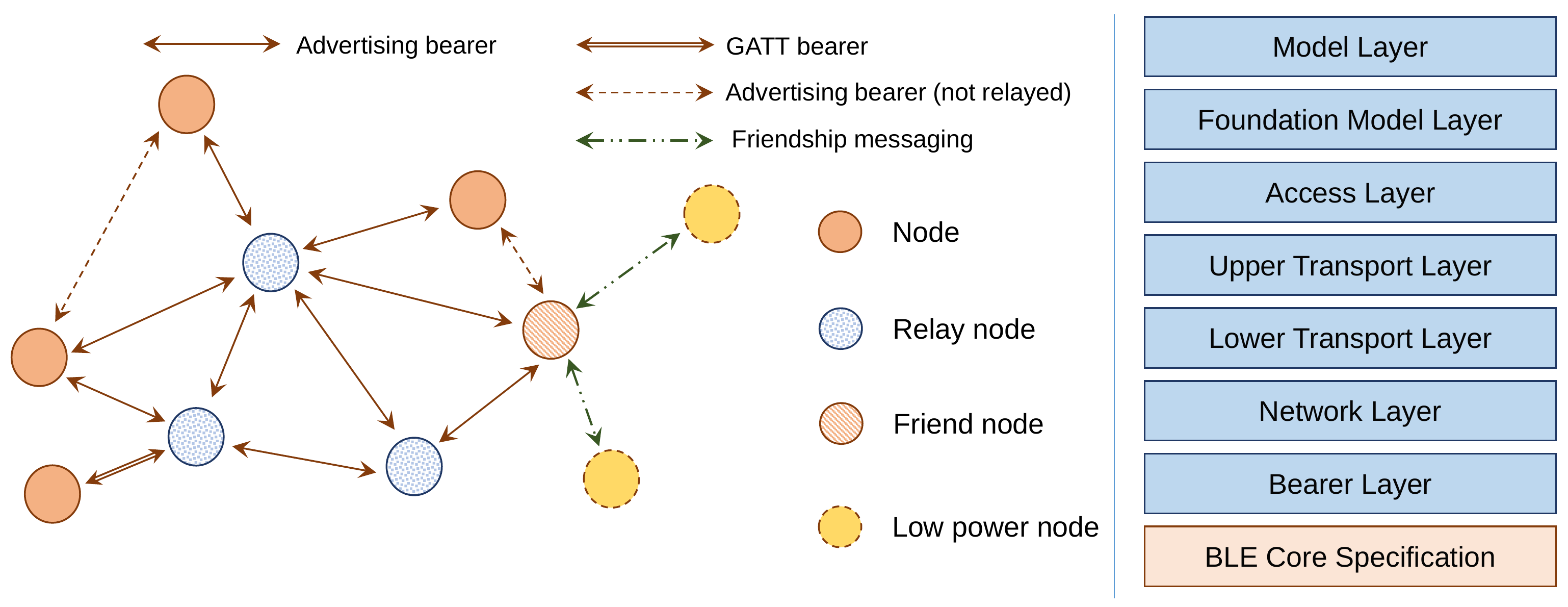}
\caption{System architecture and protocol stack of Bluetooth mesh. }
\label{arch_stack}
\vspace{-1.5em}
\end{figure}

\subsection{Related Work}
Yin \emph{et al.} \cite{survey_BT_mesh_5} conducted a comprehensive survey of Bluetooth mesh and Bluetooth 5.0. 
Rondon \emph{et al.} \cite{und_perf_BT_mesh} investigated performance of Bluetooth mesh, in terms of reliability, scalability and delay, through simulations on a grid topology. Hernandez-Solana \emph{et al.} \cite{BT_mesh_analysis} adopted an analytic approach for investigating interaction of different Bluetooth mesh parameters. Di Marco \emph{et al.} \cite{eval_BT_5_modes} conducted simulations-based evaluation of different data transfer modes of Bluetooth 5.0. 

Some studies have conducted experimental evaluation of Bluetooth mesh. Leon and Nabi \cite{exp_BT_mesh_1} evaluated packet delivery performance under a convergecast traffic pattern where all nodes periodically generate data packets for a common destination. Baert \emph{et al.} \cite{exp_BT_mesh_2} investigated latency bounds with varying number of hops and relays. Jürgens \emph{et al.} \cite{mesh_ILS} investigated application of Bluetooth mesh for indoor localization through an experimental setup. 

\subsection{Contributions}
Against this background and existing literature, this paper has two main objectives. The first is to conduct experimental evaluation of Bluetooth mesh with an emphasis on performance aspects which are not well-investigated in literature. The second is to investigate potential  enhancement/optimization techniques while providing insights into system-level performance. Our experimental evaluation aims to address the following important questions.

\begin{enumerate}

\item {What type of communication/traffic patterns are efficiently supported by Bluetooth mesh?}

\item {What is the latency performance of Bluetooth mesh for perfect reliability, under different modes of operation?}

\item {Can the performance of Bluetooth mesh be enhanced through simple techniques and parametric adjustments?}


\item {What kind of IoT applications can be supported by Bluetooth mesh?}

\end{enumerate}

We conduct experimental evaluation of Bluetooth mesh based on a testbed of Nordic nRF52840\footnote{https://www.nordicsemi.com/Software-and-Tools/Development-Kits/nRF52840-DK} devices.  The key distinguishing aspects of our evaluation include different communication patterns, different modes of operation, varying traffic loads, impact of message segmentation, and latency performance for 100\% reliability.
In terms of performance optimization of Bluetooth mesh, we investigate the role of advertising/scanning parameter adjustment, extended advertisements, dynamic power control techniques, and varying number of relays, from a system-level perspective.

\section{Overview of Bluetooth Mesh Technology}
\subsection{Protocol Stack and Architecture}
Fig. \ref{arch_stack} shows the protocol stack of Bluetooth mesh technology. Bluetooth mesh is built over the Bluetooth low energy (BLE) standard, sharing the same Physical (PHY) and Link layers.  Other layers are described as follows.

The \emph{bearer layer} handles delivery of mesh PDUs at the link layer. The default bearer is the \emph{advertising} bearer that exploits BLE advertising/scanning features to transmit/receive mesh PDUs. There is a \emph{generic attribute profile (GATT) bearer} for devices not supporting Bluetooth mesh to communicate with mesh nodes. The \emph{network layer} determines how transport messages are addressed and dealt with, and decides whether to relay, process, or reject a message. The \emph{lower transport layer} defines segmentation and reassembly process for mesh PDUs. It also provides acknowledged/unacknowledged message delivery. 
The \emph{upper transport layer} handles encryption, decryption, and authentication functions. It also handles control messages which are exchanged between peer nodes. The \emph{access layer} acts as an intermediary between the model layers and upper transport layer by providing the upper transport layer with a standard format. 
The \emph{foundation model layer} implements models related to  management of the Bluetooth mesh network. 
The \emph{model layer} defines models that provide a standardized operation for typical use-cases. Models can also be custom-defined. There are three types of models: server model, client model, and control model.

\begin{figure}
\centering
\includegraphics[scale=0.28]{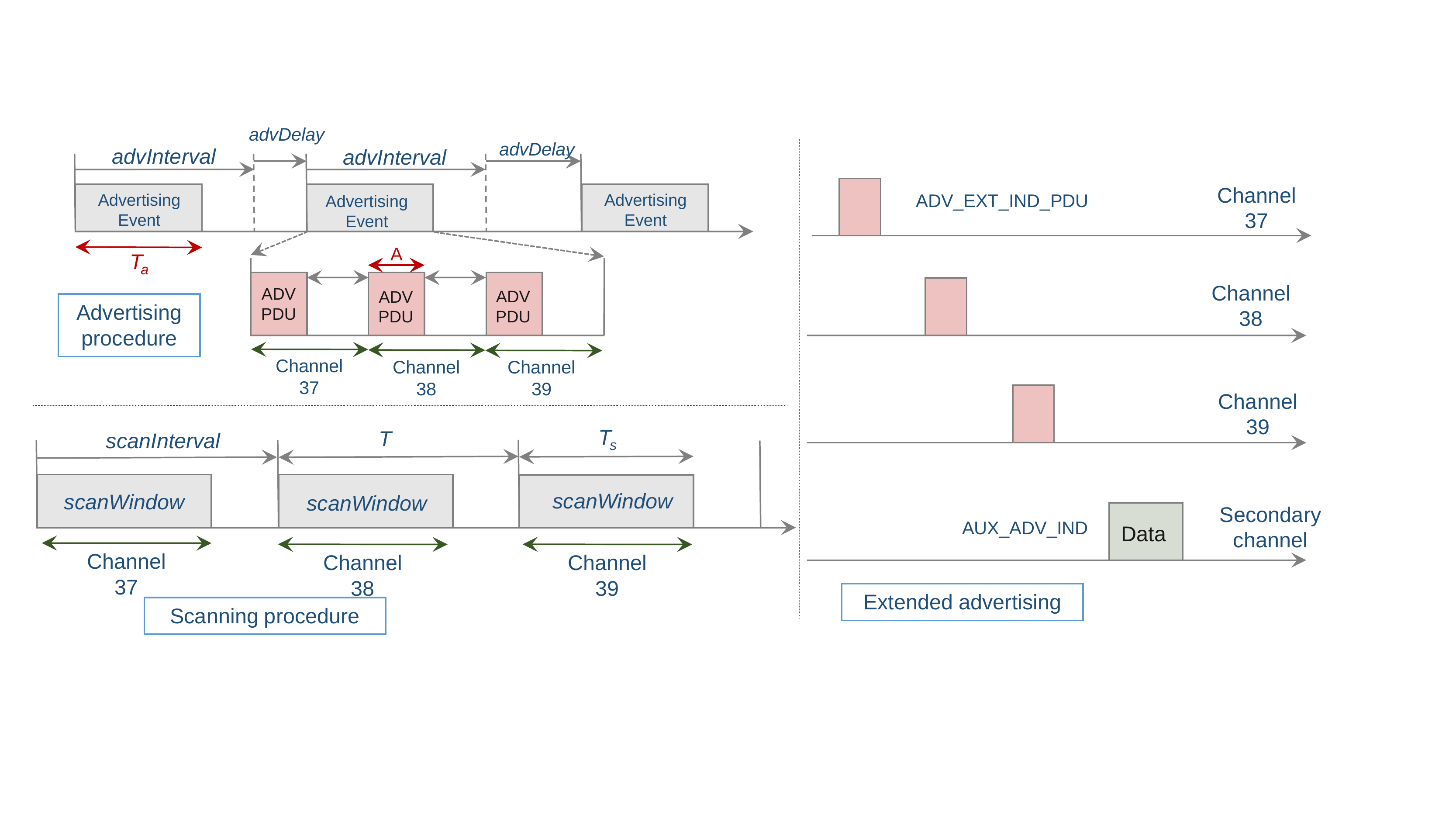}
\caption{Illustration of advertising and scanning procedures in Bluetooth mesh. }
\label{adv_scan}
\vspace{-1.5em}
\end{figure}

Fig. \ref{arch_stack} also shows the architecture of Bluetooth mesh. The term \emph{node} is used for any device which is part of the Bluetooth mesh network. A device becomes part of the mesh network through the provisioning process. All nodes in the mesh network are capable of transmitting/receiving messages; however, some optional features provide additional capabilities. 
The \emph{relay nodes} are able to retransmit received messages over the advertising bearers. Based on relaying, a message can traverse the whole multi-hop mesh network. 
A \emph{low power node} (LPN) is power-constrained and must use its energy resources as efficiently as possible. LPNs operate in a mesh network with significantly reduced duty cycle.
A \emph{friend node} assists the operation of LPNs in the mesh network by storing messages for these nodes and forwarding upon request. 

\subsection{Managed Flooding}
Bluetooth mesh utilizes a \emph{managed flooding} approach for communication. The managed flooding mechanism is completely asynchronous and provides a simple approach to propagate messages in the mesh network using broadcast. A message transmitted in the mesh network is potentially forwarded by multiple relay nodes.

Message transmissions primarily take place over the advertising bearer. Advertising is the process by which a source node injects its message in the mesh network. An advertising event is a cycle of advertising operations where mesh protocol data units (PDUs) are transmitted in sequence over each of the three (primary) advertising channels (i.e., channels 37, 38 and 39). At the network layer, multiple advertising events can be configured to improve the reliability of message injection. The time between two advertising events is dictated by the advertising interval (\emph{advInterval}) and a random advertising delay (\emph{advDelay}).
The relay nodes scan the advertising channels and listen to the advertising information of the neighbors. The scanning procedure is performed in scanning events that repeat after a scanning interval (\emph{scanInterval}). The probability of message propagation in the mesh network is increases with multiple relay nodes scanning on different advertising channels at different times. 

Managed flooding provides multi-path diversity that improves reliability. However, it can also result in increased collisions on the advertising channels. Bluetooth mesh offers various configuration options to overcome this issue, for example time-to-live (TTL) limitations, message cache to restrict forwarding, and random back-off periods between different advertising events and different transmissions within an advertising event. Bluetooth mesh implements a publish/subscribe messaging paradigm. Publishing refers to the act of sending a message. Typically, messages are sent to unicast, group or virtual addresses. 


\subsection{Extended Advertisements}
Bluetooth 5.0 introduces extended advertising that exploits additional data channels for transmission. A source node transmits short advertising indication PDUs (on primary advertising channels) that include a pointer to a secondary advertising channel (randomly selected from the other 37 BLE channels) over which data transmission takes place. Extended advertisements enable transmission of more data than that allowed on legacy advertisements. Data transmissions on secondary advertising channels can use any of the Bluetooth 5.0 PHY layers i.e., coded (500/125 kbps) and uncoded (1/2 Mbps). 

\begin{figure}
\centering
\includegraphics[scale=0.33]{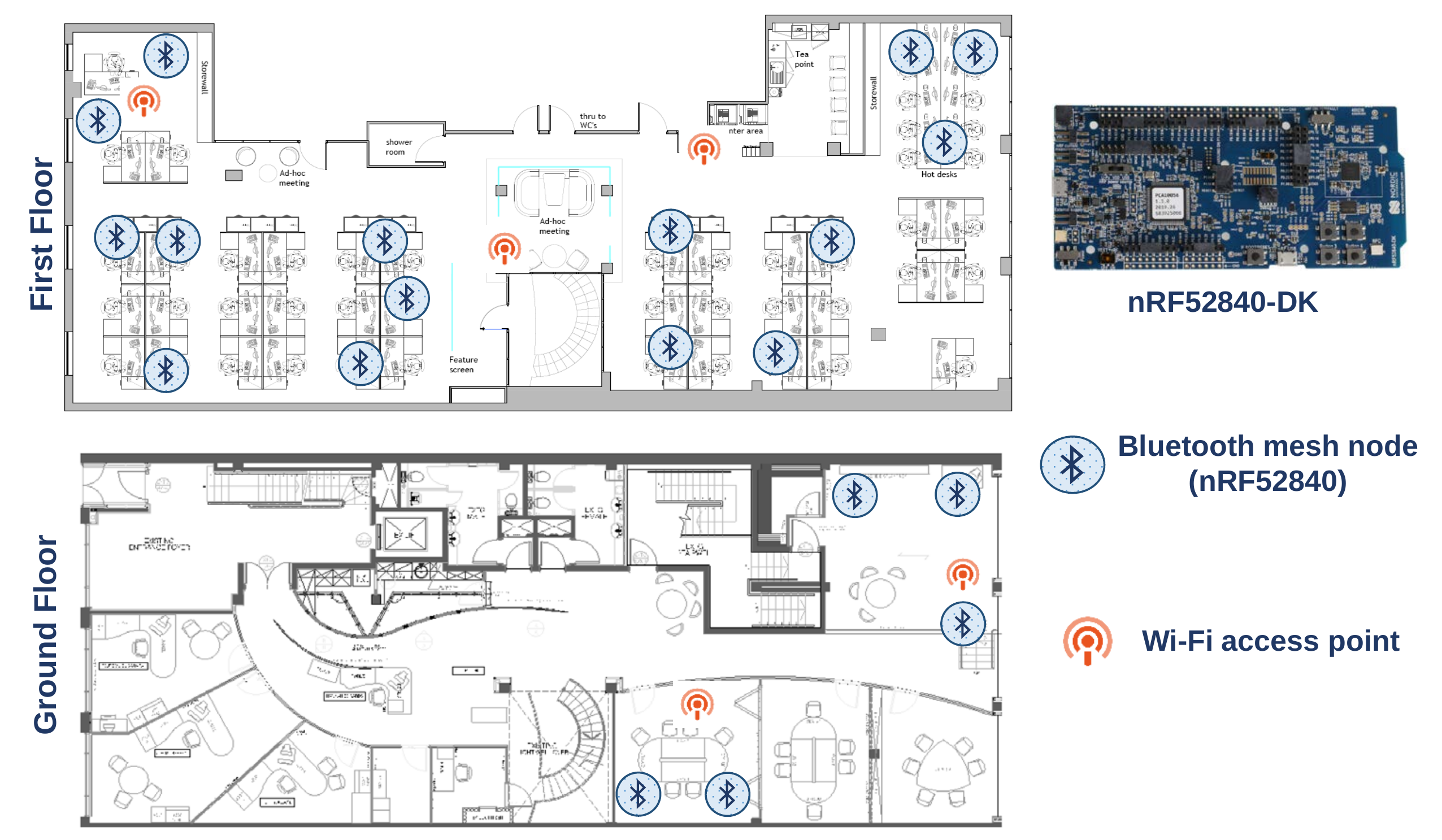}
\caption{Testbed for experimental evaluation of Bluetooth mesh.}
\label{testbed1}
\vspace{-1.5em}
\end{figure}

\section{Experimental Evaluation and Optimization}

\subsection{Testbed Setup and Evaluation Scenarios}
Fig. \ref{testbed1} shows our testbed for experimental evaluation. We have deployed 20 nRF52840 development boards over two floors of our office building covering approximately 600 square meters. The testbed setup depicts a challenging multi-hop mesh scenario due to weak link between the two floors. The multi-hop network stretches over a maximum of 4 hops.  The testbed nodes experience shadowing (from humans) and external interference (from Wi-Fi access points and other Bluetooth devices operating in office environment). We use the Nordic nRF5 software development kit (SDK) for Bluetooth mesh evaluation. We have implemented precision time protocol (PTP)\footnote{http://linuxptp.sourceforge.net/} with software timestamping to time synchronize testbed nodes for one-way latency measurements. The hex files required to flash the boards were built using SEGGER Embedded Studio for ARM v4.12. bash scripts to setup and program the boards as well as a php script for latency/reliability capture.

We consider different evaluation scenarios for Bluetooth mesh in terms of traffic patterns (one-to-many, many-to-one, and many-to-many), varying number of concurrent senders, varying message sizes with and without segmentation, and unicast and group communication modes of operation. 

The default setting of key parameters in our evaluation is given in \tablename~\ref{def_params} (unless stated otherwise). Experimental results are repeated over 100 iterations with a new message every 1000 ms. In case of many-to-many communication, the source and destination nodes are at least two hops away. 

\begin{center}
	\begin{table}
		\caption{Default parameters and configuration for evaluation}
		\begin{center}
			\begin{tabular}{ll}
				\hline	
				\toprule
				Parameter  & Value  \\\hline
				\midrule
				Transmit power   & 0 dBm \\ 
				Advertising interval (\emph{advInterval}) & 20 ms \\
				Scanning interval (\emph{scanInterval}) & 2000 ms \\
				Relay configuration & All nodes are relays \\
				No. of advertising events & 3 (source node); 2 (relay node)\\
				Message size & 11 bytes (unsegmented) \\
				Mode of operation & Unicast (acknowledged) \\
				\hline
			\end{tabular}
		\end{center}
			\label{def_params}
			\vspace{-1.5em}
	\end{table}
\end{center}

\subsection{Evaluation of Unicast and Group Communication Modes}
\begin{figure}
\centering
\includegraphics[scale=0.28]{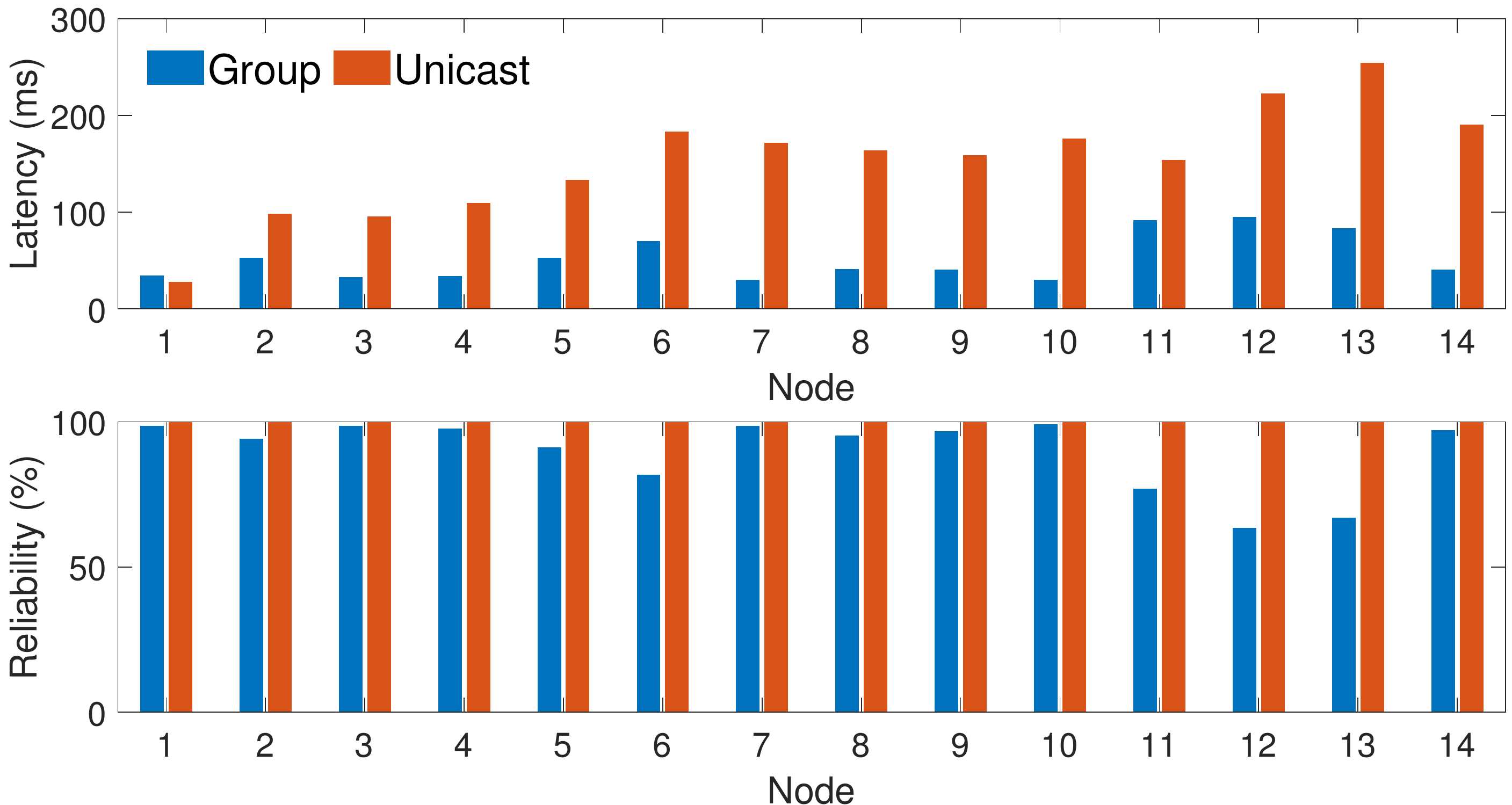}
\caption{Performance of unicast and group modes (multi-hop scenario).}
\label{UG1}
\vspace{-1 em}
\end{figure}

\begin{figure}
\centering
\includegraphics[scale=0.28]{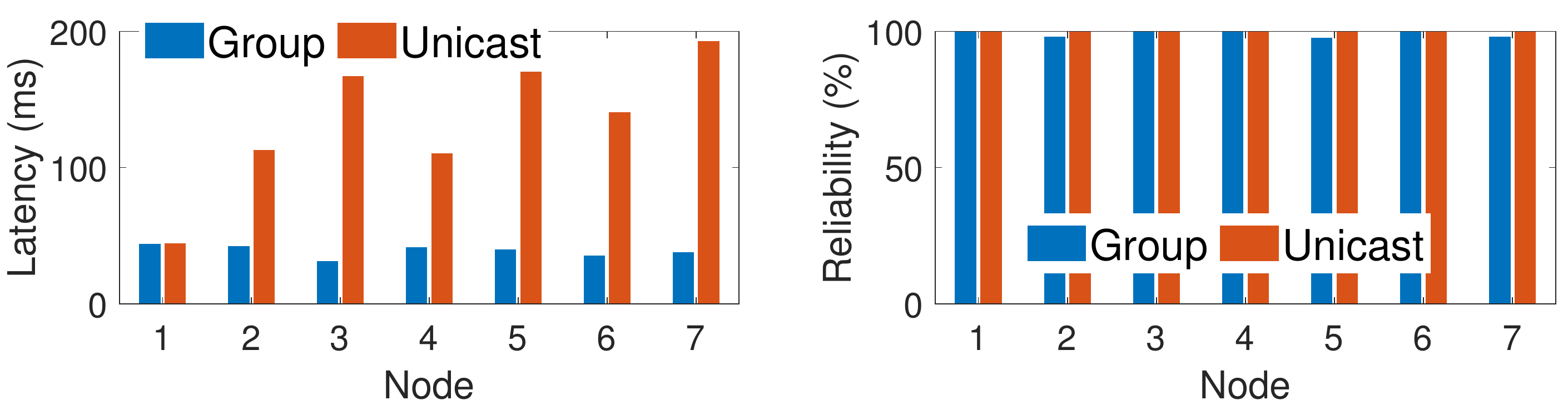}
\caption{Performance of unicast and group modes (single-hop scenario)}
\label{UG2}
\vspace{-1 em}
\end{figure}

First, we evaluate the latency and reliability performance of unicast and group communication modes. We consider one-to-many and many-to-one communication scenarios where one of the nodes is selected as a controller while others act as slaves. The controller sends a command message to each slave node which creates a one-to-many traffic pattern. Moreover, we consider acknowledged mode for mesh messages. Hence, slave nodes acknowledge command messages from the controller which creates a many-to-one traffic pattern. We define two different groups of nodes. The first group consists of 15 nodes (1 controller and 14 slaves) in a multi-hop mesh network spanning both floors. The second group comprises 8 nodes (1 controller and 7 slaves) as a single-hop network (left part of the first floor). In the unicast mode, the controller sends a message to each node individually. The message is acknowledged; hence, the controller keeps sending until an acknowledgement is received. In the group mode, the controller sends a group message to all slave nodes. This message is also acknowledged; however, it is only sent twice (i.e., over two advertising events). For both communication modes and groups, we measure the round-trip latency and the reliability (in terms of packet delivery) under default set of parameters. The performance results, in terms of round-trip latency and reliability for each node, are shown in Fig. \ref{UG1} and Fig. \ref{UG2} and summarized in \tablename~\ref{t_ug}. As shown by the results, the unicast mode achieves 100\% reliability; however, it incurs higher latency. The group mode provides significantly lower latency; however, its reliability is affected by the fixed number of message transmissions.

\begin{table}[]
\vspace{-1em}
\caption{Summary of performance in unicast and group modes}
\begin{tabular}{ccccc}
\hline
\textbf{Scenario}                               & \textbf{Mode}             & \multicolumn{1}{c}{\textbf{\begin{tabular}[c]{@{}c@{}}Mean \\ Reliability\end{tabular}}} & \multicolumn{1}{c}{\textbf{\begin{tabular}[c]{@{}c@{}}Mean Latency \\ (round-trip)\end{tabular}}} & \multicolumn{1}{c}{\textbf{\begin{tabular}[c]{@{}c@{}}Max. Latency \\ (round-trip)\end{tabular}}} \\ \hline \hline
\multirow{2}{*}{Multi-hop}                      & Unicast                   & \multicolumn{1}{c}{100\%}                                                                & \multicolumn{1}{c}{152.27 ms}                                                                     & \multicolumn{1}{c}{253.62 ms}                                                                     \\ 
                                                & \multicolumn{1}{l}{Group} & 89.6\%                                                                                   & 51.62 ms                                                                                          & 94.37 ms                                                                                          \\ \hline
\multicolumn{1}{l}{\multirow{2}{*}{Single-hop}} & Unicast                   & 100\%                                                                                    & 133.77 ms                                                                                         & 192.57 ms                                                                                         \\ 
\multicolumn{1}{l}{}                            & \multicolumn{1}{l}{Group} & 99.07\%                                                                                  & 38.57 ms                                                                                          & 43.45 ms                                                                                          \\ \hline
\end{tabular}
\label{t_ug}
\vspace{-1.5em}
\end{table}

\subsection{Latency Performance for 100\% Reliability}
\begin{figure}
\centering
\includegraphics[scale=0.3]{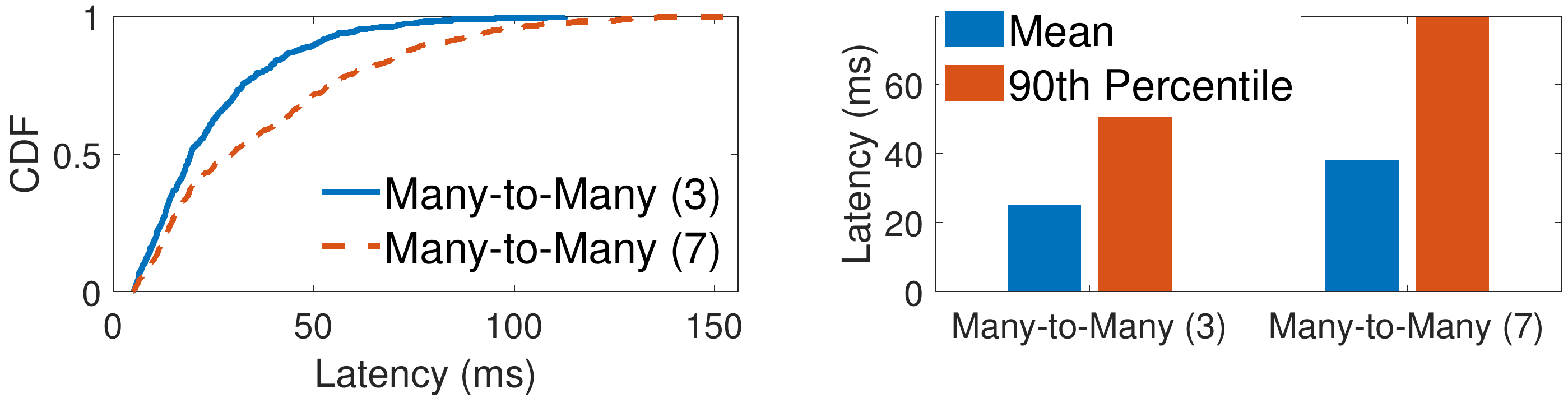}
\caption{Latency performance for 100\% reliability (default parameters).}
\label{lat_def}
\vspace{-1 em}
\end{figure}

Next, we evaluate the latency performance for 100\% reliability (which is given by the unicast mode) under default set of parameters. We consider a many-to-many multi-hop communication scenario wherein either 3 concurrent senders transmit to a group of 3 distinct destinations or 7 concurrent senders transmit to a group of 7 distinct destinations. The 3 concurrent senders' scenario represents a low to medium traffic scenario whereas the 7 concurrent senders' scenario represents a high traffic scenario. The senders and receivers are randomly selected from the entire multi-hop mesh network in each iteration. The results for one-way latency measurement are shown in Fig. \ref{lat_def}.
The mean latency for many-to-many scenario with 3 concurrent senders is 24.94 ms and the 90th percentile is 50.41 ms. With 7 concurrent senders, the mean latency increases to 37.89 ms and the 90th percentile is 79.59 ms. The higher latency is due to more traffic on the advertising channels which leads to more retransmissions.

\subsection{Impact of Message Segmentation}
\begin{figure}
\centering
\includegraphics[scale=0.3]{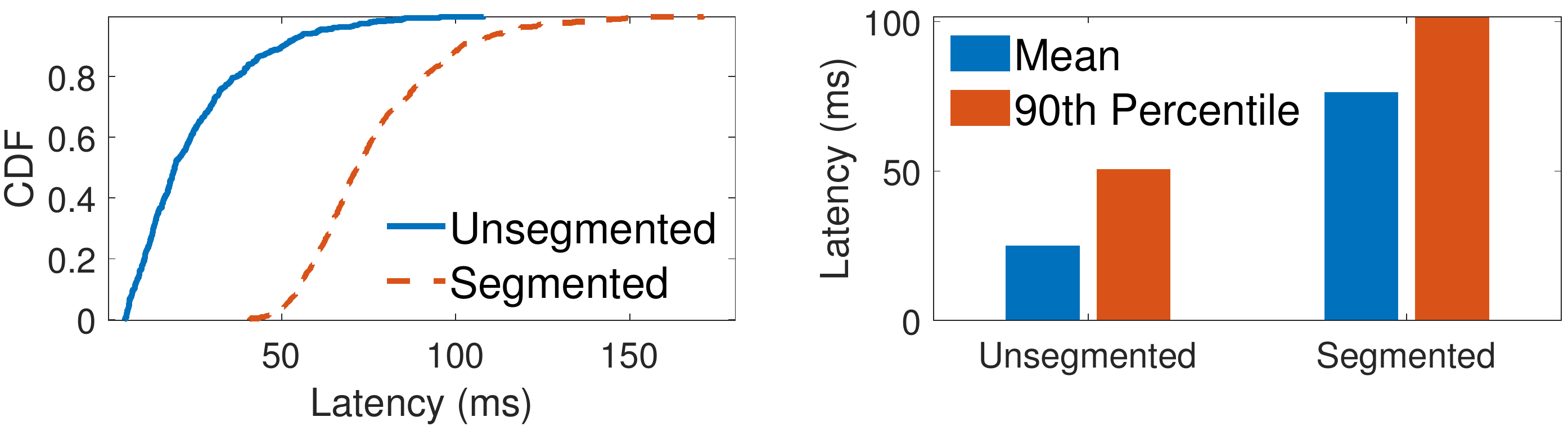}
\caption{Latency performance of segmented versus unsegmented messages.}
\label{seg}
\vspace{-1 em}
\end{figure}

The lower transport layer applies a segmentation and reassembly procedure for messages above 11 bytes. Hence, it is important to evaluate the impact of message segmentation. Fig. \ref{seg} shows one-way latency performance for segmented and unsegmented messages in case of many-to-many communication with 3 concurrent senders. We have used 11 byte and 19 byte messages for unsegmented and segmented scenarios, respectively. The mean latency for unsegmented messages is 24.94 ms and the 90th percentile of 50.41 ms. The mean latency for segmented messages increases more than twice to 76.32 ms and the 90th percentile is 101.67 ms. The results reveal that segmentation process incurs higher latency as more messages are injected in the mesh network.


\subsection{Performance Optimization with Parameter Adjustments}
\begin{figure}
\centering
\includegraphics[scale=0.3]{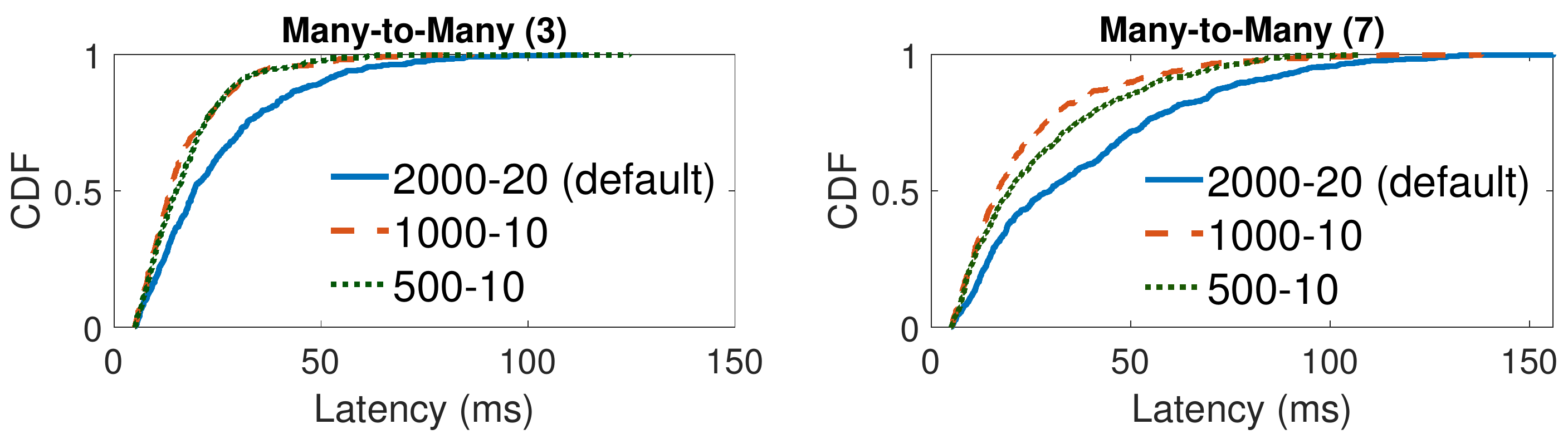}
\caption{Latency performance with different scanning/advertising parameters.}
\label{lat_param}
\vspace{-1 em}
\end{figure}

Next, we investigate performance optimization of Bluetooth mesh by simple adjustment of two key parameters:  \emph{advInterval} and \emph{scanInterval}. Both parameters must be adjusted in tandem so that advertising and scanning procedures are affected simultaneously. We consider three different combinations of (\emph{scanInterval})-(\emph{advInterval}): 2000-20 ms (default), 1000-10 ms, and 500-10 ms. The one-way latency performance for many-to-many communication with 3 and 7 concurrent senders is shown in Fig. \ref{lat_param} and further summarized in \tablename~\ref{t_param_adj}. The results show that default advertising/scanning parameters do not provide the best latency performance. Compared to 2000-20 ms, 1000-10 ms combination reduces the mean latency by nearly 31.5\% and 38.4\% for the cases of 3 and 7 senders, respectively. The reduction in latency is due to relatively faster advertising of messages. The 500-10 ms combination reduces latency compared to the default case; however, it increases  latency compared to 1000-10 ms combination as the relays spend relatively less time scanning the advertising channels. 

\begin{table}[]
\caption{Summary of performance with parameter adjustments.}
\begin{tabular}{cccc}
\hline
\textbf{Scenario}                                     & \multicolumn{1}{c}{\textbf{\begin{tabular}[c]{@{}c@{}}Parameter \\ Combination \end{tabular}}} & \multicolumn{1}{c}{\textbf{\begin{tabular}[c]{@{}c@{}}Latency \\ (Mean)\end{tabular}}} & \multicolumn{1}{c}{\textbf{\begin{tabular}[c]{@{}c@{}}Latency \\ (90th percentile)\end{tabular}}} \\ \hline \hline
\multirow{3}{*}{Many-to-Many (3)}                                       & 2000-20 (ms) & 24.94 ms & 50.41 ms \\
                                                 & 1000-10 (ms) & 17.09 ms                            & 31.18 ms\\    
                                                 & 500-10 (ms)                                                                                          & 23.22 ms                            & 50.69 ms                                                           \\ \hline
\multicolumn{1}{l}{\multirow{3}{*}{Many-to-Many (7)}}                   & 2000-20 (ms)                                                                                    & 37.89 ms                                                                                         & 79.59 ms \\ 
\multicolumn{1}{l}{}                            & 1000-10 (ms)  & 23.22 ms                                                                                          & 50.7 ms     \\
\multicolumn{1}{l}{}   &500-10 (ms)                                                                                          &26.69 ms &56.5 ms \\
 \hline
\end{tabular}
\label{t_param_adj}
\end{table}

\subsection{Performance Optimization with Extended Advertisements}
Extended advertisements are promising for latency reduction due to reduced contention on primary advertising channels. Moreover, extended advertisements can exploit the 2 Mbps PHY layer on secondary advertising channels that reduces transmission time compared to the 1 Mbps PHY layer for primary advertising channels. 
Since the next version of Bluetooth mesh specification with extended advertisements in still in progress, we use the Nordic proprietary Instaburst (https://infocenter.nordicsemi.com/index.jsp) feature for evaluation. Instaburst uses a subset of Bluetooth 5.0 extended advertisements with 2 Mbps PHY layer. When instaburst is enabled, all communication in the Bluetooth mesh network takes place via extended advertisements. 

\begin{figure}
\centering
\includegraphics[scale=0.3]{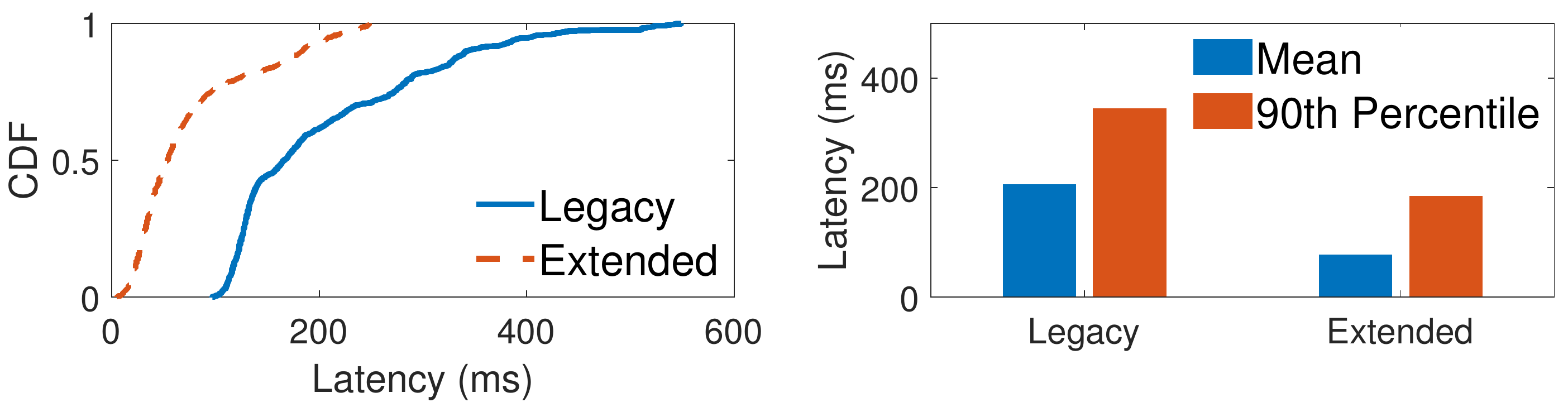}
\caption{Latency performance with extended advertisements (one-to-many).}
\label{ext1}
\vspace{-1 em}
\end{figure}

\begin{figure}
\centering
\includegraphics[scale=0.3]{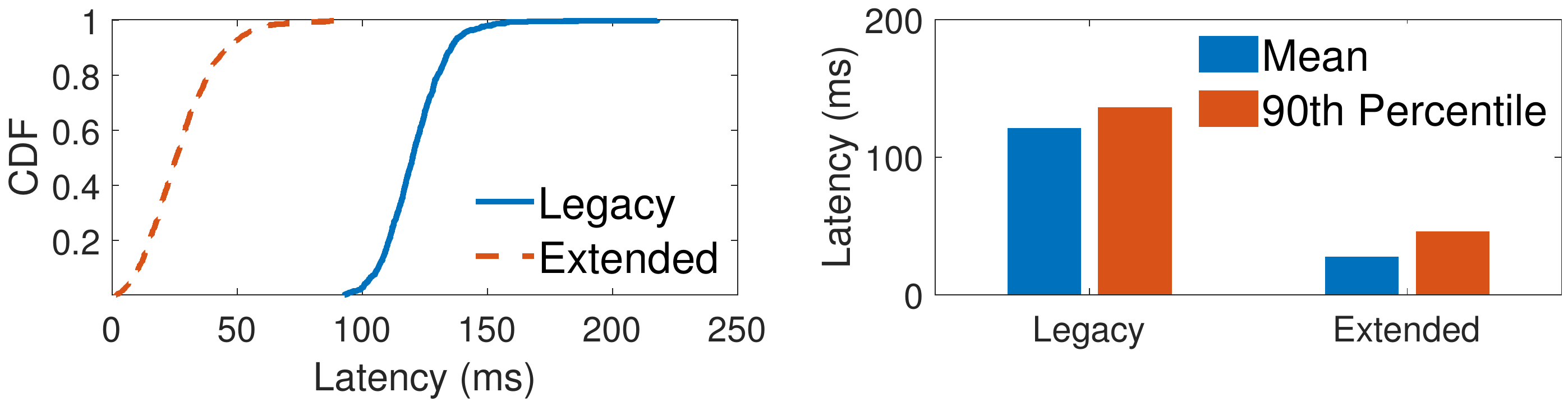}
\caption{Latency performance with extended advertisements (many-to-many).}
\label{ext2}
\vspace{-1 em}
\end{figure}

We evaluate the latency of legacy and extended advertisements for 100\% reliability under default parameters; however, with larger 50 byte messages. First, we evaluate the round-trip latency of one-to-many communication with 15 nodes. The results (shown in Fig. \ref{ext1}) reveal that the mean latency of legacy advertisements is 204.88 ms with 90th percentile of 343.5 ms. With extended advertisements, the mean latency is 76.35 ms with 90th percentile of 184 ms. Next, we evaluate one-way latency of many-to-many communication with 3 concurrent senders. The results (shown in Fig. \ref{ext2}) reveal that the mean latency of legacy advertisements is 120.8 ms with 90th percentile of 135.72 ms. With extended advertisements, the mean latency is 27.28 ms with 90th percentile of 45.67 ms. The results clearly highlight the effectiveness of extended advertisements for latency reduction of Bluetooth mesh. Specifically, the mean round-trip latency for one-to-many communication as well as the mean one-way latency for many-to-many communication reduces by up to 62\%.

%

\subsection{Performance Optimization with Power Control}
Power control is a well-known concept in wireless networks. Implementing power control in a Bluetooth mesh network provides two key benefits. The first is reduced interference to other source nodes which enables multiple transmissions at the same time. The second is higher availability of relay nodes, particularly in dense deployments, for forwarding messages from different source nodes due to reduced transmit power. 

We implement a simple and widely used power control strategy wherein a source/relay node computes an optimized power level based on the received signal strength such that

\begin{equation}
\label{eq_pc}
P_{ctl}=P_{max}\cdot(P_r)^{-1}\cdot\zeta^{th}\cdot {c},
\end{equation}
where $P_{ctl}$ is the optimized transmit power, $P_{max}$ the maximum transmit power, $P_r$ is the minimum received power, $\zeta^{th}$ denotes the minimum required received
signal strength, and $c$ is a constant \cite{pwr_control1, pwr_control2}. The minimum received power is computed based on passive listening of the advertising channels. The power control strategy is implemented only if a node has overheard messages over all advertising channels. 

\begin{figure}
\centering
\includegraphics[scale=0.3]{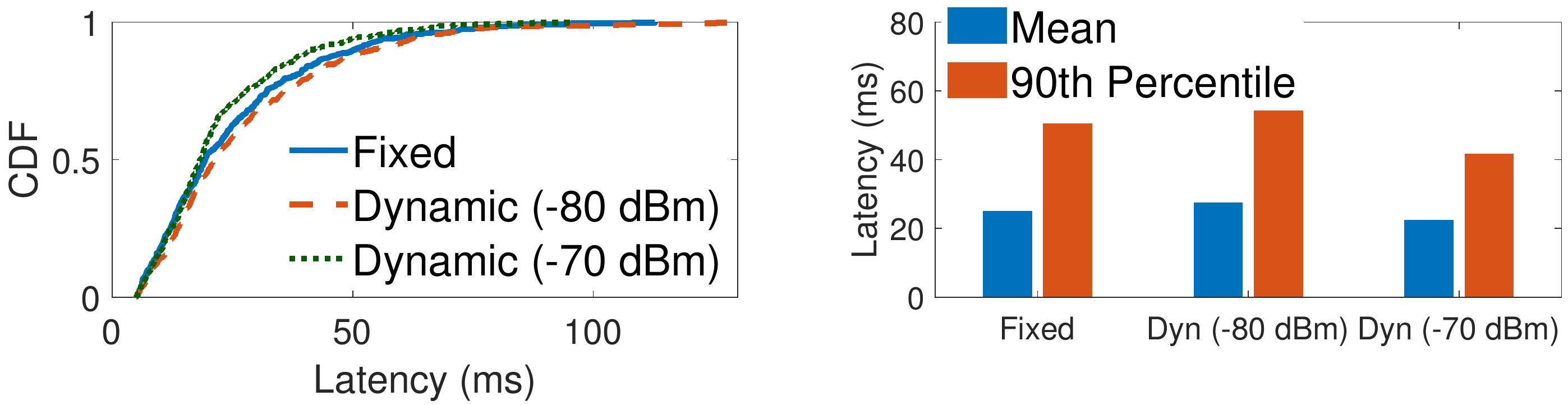}
\caption{Latency performance with power control (many-to-many with 3 senders).}
\label{pwc}
\end{figure}

We have evaluated the impact of dynamic power control under default parameters for many-to-many communication with 3 concurrent senders and with two different values of \(\zeta^{th}\): -80 dBm and -70 dBm. The results are shown in Fig. \ref{pwc}. The mean one-way latency for fixed power (i.e., no power control) is 24.94 ms with 90th percentile of 50.41 ms. For \(\zeta^{th}\) of -80 dBm, the mean latency increases by 10\% to 27.43 ms and the 90th percentile is 54.26 ms. For \(\zeta^{th}\) of -70 dBm, the mean latency decreases by 10.5\% and the 90th percentile is 41.67 ms. The increased latency for \(\zeta^{th}\) = -80 dBm is due to more aggressive power control which potentially increases link level failures leading to more message retransmissions. The latency reduction for \(\zeta^{th}\) = -70 dBm is due to more optimized power control which not only reduces interference but also maximizes availability of relays for message forwarding from multiple source nodes. The results reveal that \(\zeta^{th}\) must be carefully selected depending on network density. 


\subsection{Performance Optimization with Customized Relaying}
\begin{figure}
\centering
\includegraphics[scale=0.3]{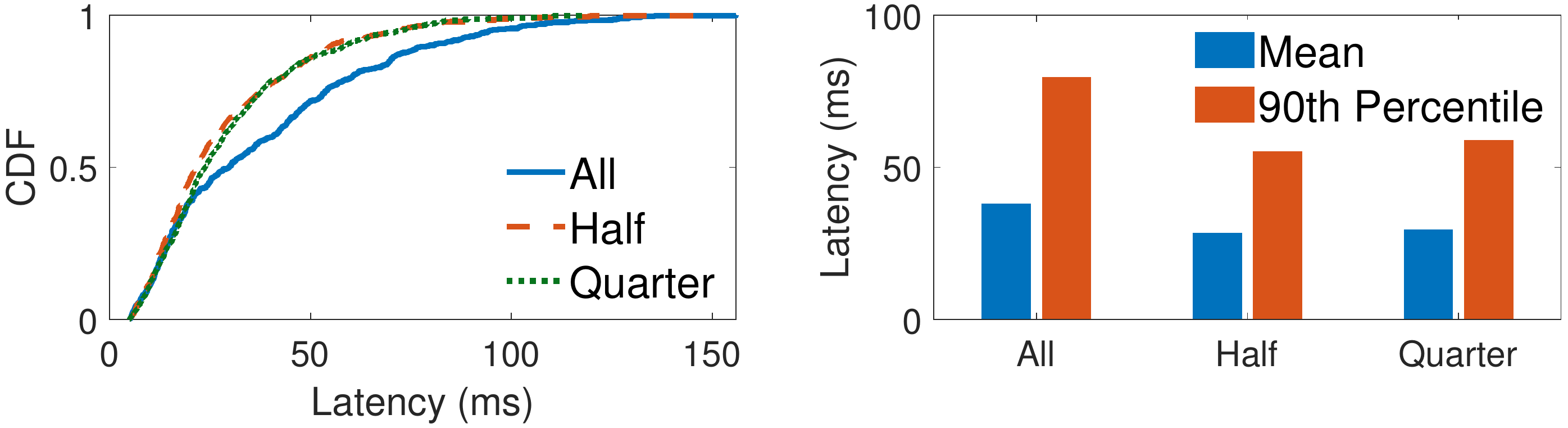}
\caption{Latency performance of customized relaying (many-to-many with 7 senders).}
\label{relay}
\end{figure}

Finally, we evaluate the impact of the number of relay nodes in a Bluetooth mesh network. By default all nodes in our network are configured as relays. For comparison, we define two additional scenarios: (a) half of the network is configured as relays, and (b) quarter of the network is configured as relays. In both scenarios, the relays are randomly selected. We consider many-to-many communication with 7 concurrent senders to evaluate the impact of relaying. The one-way latency performance for 100\% reliability are shown in Fig. \ref{relay}. The mean latency for the default case of whole network as relay is 37.89 ms with  90th percentile of 79.59 ms. With half of the network configured as relays, the mean latency reduces by 25\% to 28.37 ms and the 90th percentile is 55.13 ms. With quarter of the network configured as relays, the mean latency reduces by 22.5\% to 29.35 ms and the 90th percentile is 58.83 ms. The higher latency of quarter relays compared to half relays is due to relatively lower availability of relays for source nodes. Nevertheless, the results reveal that appropriate number of relays for a given network density plays an important role in latency reduction.


\section{Key Insights}
Our experimental evaluation of Bluetooth mesh and its performance optimization provides the following key insights. 

\begin{itemize}
\item Bluetooth mesh can efficiently support different communication patterns in unicast as well as in group modes. 

\item Bluetooth mesh can achieve perfect reliability even under high and frequent traffic; however, default parameters do not provide optimized latency performance. 

\item The performance of Bluetooth mesh can be optimized through (a) simple adjustment of parameters related to advertising/scanning, (b) simple power control techniques, and (c) selecting appropriate number of relay nodes for a given network density. 

\item Extended advertisements are promising for latency reduction, particularly for transmitting larger messages. 

\item The capability of Bluetooth mesh to support concurrent multicast and potential latency optimizations make it highly scalable. 

\item The design flexibility of Bluetooth mesh and its ability to support bi-directional information exchange in unicast as well as group communication modes under diverse traffic patterns makes it suitable for versatile control-centric and monitoring-centric IoT applications.  

\end{itemize}

The observed performance enhancement through adjustment of advertising/scanning parameters builds a strong case for adaptive advertising/scanning techniques (such as those proposed in \cite{immune_patent}). Moreover, dynamic switching between extended and legacy advertisements for mixed message sizes is promising for further latency optimization. Evaluation of such aspects will be the focus of our future work.

\section{Concluding Remarks}
The addition of mesh networking capabilities to the ubiquitous Bluetooth technology is promising for IoT applications. We have conducted experimental evaluation of Bluetooth mesh based on a testbed of Nordic nRF52840 devices in a real world environment. Our performance evaluation not only fills gaps in state-of-the-art studies but also clarifies various issues highlighted in literature. Moreover, it provides a number of insights in system-level performance of Bluetooth mesh. In particular, it reveals that Bluetooth mesh can efficiently handle different communication patterns (one-to-many, many-to-one, and many-to-many) and varying traffic loads in unicast as well as group modes. The latency performance for perfect reliability can be optimized through adjustment of advertising/scanning parameters (reducing latency by more than 30\%), extended advertisements (reducing latency by more than 60\%), simple power control techniques (providing up to 10\% latency reduction), and customized relaying (providing more than 20\% latency reduction). Bluetooth mesh provides a flexible solution which can be applied to various monitoring and control applications. 
The managed flooding approach also ensures transparency to underlying network topology; hence Bluetooth mesh also can be applied to mobility-centric applications.

\bibliographystyle{IEEEtran}
\bibliography{mesh.bib}
\end{document}